\documentclass{easychair}

\usepackage{doc}
\usepackage{wrapfig}

\title{Static analysis tools in the era of cloud-native systems}

\author{
Tomas Cerny
\and
Davide Taibi
}

\institute{
  Computer Science, Baylor University,
  Waco, Texas, USA\\
  \email{tomas\_cerny@baylor.edu}
\and
   Tampere University,
   Tampere, Finland, EU\\
   \email{davide.taibi@tuni.fi}
 }

\authorrunning{Cerny and Taibi}

\titlerunning{Static analysis tools in the era of cloud-native systems}

\begin{document}

\maketitle

\begin{abstract}

Microservices fuel cloud-native systems with small service sets developed and deployed independently. The independent nature of this modular architecture also leads to challen-ges and gaps. The intended system design might deviate far from what is eventually produced and maintained as the architecture tends to degrade over time.
% . The never-ending system evolution, may degrade the system architecture and introduce various deficiencies impacting system performance, maintenance, or management costs.  
%   Cloud-native systems bring many outstanding benefits. They provide system decentralization, scalability, and flexible evolvability needed for dynamic and agile environments. These systems solve enterprise problems, and one cannot expect a small team to be responsible for the full stack responsibilities. It is inevitable for new challenges to emerge in various areas related to multi-team involvement, decentralized evolution, and separate codebases for system parts empowered by microservices. Sooner or later, without a centralized governance of microservices, the architecture might start degrading, leaving one in void about process handling, compliance of policy enforcement, consistency, etc. 
This paper challenges the audience on how static analysis could contribute to microservice system development and management, particularly managing architectural degradation. It elaborates on challenges and needed changes in the traditional code analysis to better fit these systems and discusses implications for practitioners once robust static analysis tools~become~available. 
% . As an example, we point out the missing system-centric perspective of cloud-native systems, which we know can be derived from the codebases of these systems. We discuss its implications on practitioners when robust static analysis tools become available. 
\end{abstract}

% The table of contents below is added for your convenience. Please do not use
% the table of contents if you are preparing your paper for publication in the
% EPiC Series or Kalpa Publications series

% \setcounter{tocdepth}{2}
% {\small
% \tableofcontents}

%\section{To mention}
%
%Processing in EasyChair - number of pages.
%
%Examples of how EasyChair processes papers. Caveats (replacement of EC
%class, errors).

%------------------------------------------------------------------------------
\section{Introduction}
\label{sect:introduction}

% Cloud-native systems are designed to take full advantage of the cloud infrastructure.
% % The 12-factor app methodology~\cite{TheTwelv82:online} provides guidance on designing, building, and deploying these systems. For instance, 
% Cloud-native systems have, as the foundation the microservice architecture, utilize containers, and follow Continuous Integration and Delivery (CD/CI). 

Microservice architectural style is a paradigm to develop systems as a suite of small, self-contained, and autonomous services, communicating through a lightweight protocol.
Each microservice has own codebase with a separated configuration to facilitate evolution. This, as a result, enables the separation of duty for roles like architects, developers, and DevOps.

What is, however, less perfect is the separation of concerns. Such separation is likely to be well managed on a single codebase level, but it might get lost with the decentralization and existence of multiple codebases. 
% While infrastructure like centralized configuration servers and API gateway exists, i.e., to enable telemetry and tracing, these cannot be misused beyond their original purpose. 

% Another factor that must be highlighted is the existence of certain overlaps across microservices with their bounded contexts. This 

Overlaps across microservices with their bounded contexts are inevitable as microservices interact. Implications from these overlaps are reflected in the microservice codebase. Since microservice codebase remains self-contained, overlaps means partially restated definitions, typically reimplemented in a particular framework version. This restatement can relate to data definitions of processed information, encapsulated knowledge, business logic, or other enforcement related to various policies (i.e., security, constraints, privacy, etc.).
However, this overlap is uncontrolled throughout the system evolution and there are fragile mechanisms to assess consistency errors. Thus, once any of these definitions changes in the microservice codebase, there is no direct indication of the definition being restated elsewhere in other codebases.

% If we reverse the separation of concern perspective. 
We typically want to separate concerns to provide better readability and maintainability. We can do a micro-management and design solid concern separation per each microservice. However, this would only relate to a single microservice and not the whole system. The question to ask is whether we  need to see a certain concern per the entire system perspective. Suppose we are architects, most likely yes. Even for developers, it would be nice to see concerns aligned across interacting microservices. However, each concern of a certain type is re-defined and encapsulated across microservices. This might be one of the greatest disappointments when migrating from monolith systems. As an example, the consequence for security assessment is that each microservice has to be analyzed individually, and then extracted knowledge must be combined ad-hoc, which is tedious, error-prone, and does not scale with agile development. 

In this work, we discuss how static analysis could contribute to solve the shortcomings of microservices-based systems. We emphasize how the future tools should adapt to better fit these systems' specifics. We base our discussion on a set of prototype tools that we developed with our research teams. 

In the remainder of this paper, we discuss the current approaches to assess cloud-native systems (Section 2). Next, Section 3 focuses on changes to static analysis tools to better align with cloud-native. Finally, Section 4 discusses the implications and impact on involved stakeholders once these tools become robust and available, while Section 5 concludes the paper.

\section{Current Trends}
Researchers often point to dynamic system analysis as one possible direction to address decentralization challenges. Dynamic analysis, typically driven by telemetry (i.e. \url{https://opentelemetry.io}) is often considered as the answer to the system-centric perspective, derived from traces of user interaction or simulated tests. The great advantage of this approach is its platform agnosticism. Still, many additional steps must be done for microservices to be integrated and support this~\cite{carnell2021spring}. For instance, correlation ID must be introduced, log centralization must be in place, and health checks must be provided for most advanced reporting. The dynamic analysis led by telemetry can determine microservice dependencies from call-graphs~\cite{Esparrachiari:2018:TCM:3277539.3277541,Thalheim:2017:SAI:3135974.3135977,service_dependency}, a heat map of how often are certain endpoints reached. 
However, dynamic analysis does not have access to details only available in codebases. Besides, we must consider the separation of duty relevant to telemetry. It is not developers who manage telemetry, but DevOps which introduces indirection, multi-step interpretation, and latency between what has been developed and what has been found consequently from the operating system. The argumentation could be similar to whether we should use typed-safe or interpreted languages with no type-safety. Developers typically take advantage of quick code and workspace checks that are based on static analysis. These are often part of their integrated development environments, build files, or added to the CD/CI pipelines. However, these tools only relate to a single codebase. The emerging challenge is that successful new tools have to operate across codebases and combine results with seeing the system as a whole rather than as separated pieces of the puzzle.

When comparing static and dynamic analysis, we must understand that these have two different targets. One can tell us what the underlying structures and the white-box view are; the other gives us detail about how the system is used and performs and provides the black-box system overview. Both approaches can identify entry points to the system or to its microservices, and that is where we can see the overlap. However, it is also the boundary of where the approach limits stand. Anything below entry points can be assessed by static analysis, and whatever happens above is the target for dynamic analysis. However, we see that static analysis is rather in control of developers and the dynamic analysis is more relevant to DevOps. Still, for another stakeholder, i.e., to perform a security assessment, we might need a combination of both. Above this, the system can also utilize a hybrid approach to understand system insights and dynamics. Such an approach involves code instrumentation which adds additional logging or measures into the executed code. However, this approach has a performance impact and is often applied in profiling and performance optimization.

% Dynamic analysis can give us the insight of system runtime, thus system behavior in time. However, since collected traces occupy a lot of space, there are typically maintained over a limited time. Therefore we often analyze what happened with the system runtime recently. 

% The perspective of time can also be considered with static analysis. For instance, 
Mining Software Repositories (MSR) can indicate how the system structure changes over time. We can collect additional information related to version control messages, possibly linked to issues in ticketing systems. However, we must also assume one important thing; the static analysis does not only consume code or code changes. The cloud-native design typically involves build files and container configuration files in the repository, and these files can be easily analyzed to help determine topology~\cite{recovering_architecture,spe.2974,attack_graph} and involved technology.

Still, the primary input for static analysis is the system code. It is typically parsed into an Abstract-Syntax Tree (AST). Various tools then analyze the tree to perform defined verification or match patterns. The AST can also be used to generate an intermediate representation (IR) or a model in which the system information is reasoned.

\section{Static Analysis for Microservices-Based systems}

Conventional static analysis performs on a single codebase. It determines dependencies across various internal structures with a central focal point. 
% This general expectation may exist across practitioners and remain the same for cloud-native systems. 
However, cloud-native systems are decentralized with a self-contained codebase per microservice. This difference makes it more challenging to deliver anticipated results since each codebase could employ a different framework, platform, or library version. Thus, we must consider static analysis per each codebase. 

Multi-codebase is not the only challenge; we cannot just align the analysis results linearly next to each other. Instead, we need to interweave them in the scope where they overlap - across bounded contexts. If we accomplish this, we can derive a virtual holistic perspective of the system with fine granularity of system internal dependencies.

To overcome the above challenges, we need good tactics. Since many platforms can be used, it is unavoidable to employ multiple platform parsers. The result of all such efforts should be in a unified IR. This will also enable IR interweaving that does not need to deal with heterogeneity. 

In our research and prototyping\cite{walker2021automatic,closer2022vincent}\footnote{\url{https://cloudhubs.ecs.baylor.edu/prophet/}} and \cite{Imranur2019}\footnote{MicroDepGraph \url{https://github.com/clowee/MicroDepGraph}}, we focused on microservice middleware, on the detection communication patterns between services~\cite{Pigazzini2020, Taibi2019closer19,Taibi2020CLOSER} and on metrics to detect coupling based on the interaction between microservices detected with static analsyis~\cite{Panichella2021}. 

Furthermore, we observed that most microservices would be developed using particular platform frameworks that introduce components \cite{CernyIEEEAccess2020,Schiewe2022}. Among examples, consider Spring or Java Enterprise. Even if components are not employed, a good programming convention will be established following the separation of concerns on the codebase level. With a focus on such practice, we determined that low-level code analysis can be unnecessary. Instead, we can focus on components like data entities, repositories, services, and controllers. In addition, the internal call-graphs and involved high-level structures should be detected (i.e., remote-procedure calls, REST-call, event registration, etc.). With this priority, we can use AST to detect the structure and determine an intermediate representation from these interconnected components augmented with methods, properties, and additional details. We have as assessed this approach on Spring and Java Enterprise on two system benchmarks~\cite{trainticket,testbed}. 

In our follow-up work \cite{Schiewe2022}\footnote{\url{https://github.com/cloudhubs/source-code-parser}}, we intended for generalization and proposed that the AST is extended to be a superset across languages, which we call Language Agnostic AST (LAAST). Using LAAST, it is fairly simple to build or customize pattern matching agents to detect components or higher-level structures. Thus, a common set can be established for conventional framework components. Still, the developer can customize these matchers for naming conventions and apply custom callback to populate the IR with a given component. We have tested this prototype success on the previous testbeds~\cite{trainticket,testbed} and added one in C++~\cite{DeathStarBench}, manually validating precision and recall for component detection above 95\% \cite{Schiewe2022}. Still, more experiments for other platforms remain.

With the IR of each microservice, we need to interweave them. The bottom-up approach is to join involved data models. The horizontal approach is to predict possible inter-service communication. This can be accomplished by detecting remote calls to certain endpoints identifying relative paths, HTTP types, and parameters, and matching them to endpoint signatures. We recognize that only dynamic analysis can recognize inter-service communication with perfect precision, but if we solely consider static analysis, this results with the best approximation. However, other interactions based on events and brokers can be taken into account as well.

% We have applied the two ingredients
To interweave microservices, first, we identify overlap with data entities. Using LAAST matching agents we identify entities. We analyze these entities to derive a data model per each bounded context. Next, we use similarity from natural-language processing, Wu-Palmer algorithm~\cite{han-etal-2013-umbc} to determine potential matches in entities across microservice IRs. Placing identified entities in an overlay helps us connect IRs together and build a context map. However, other techniques can be used. 
Next, we consider possible remote calls between microservices. Similarly, we operate on LAAST to identify endpoints, relative paths, HTTP types and parameters, and similarly remote calls within services, which we match based on HTTP types, relative paths, and parameters. We can determine additional dependencies across microservices that strengthen the previously assembled overlay with this route. Performing this across all microservices we determine the holistic system IR, which corresponds to the latest state of the system. We have tested this interweaving on the previously mentioned testbeds~\cite{trainticket,testbed,DeathStarBench} and assessed the results manually, with few associations missing in the resulting context map and few unidentified connections in the inter-service interaction~\cite{closer2022vincent}. The missing connections were all due to ambiguity caused by choosing from multiple potential URLs at endpoints which we did not design to our prototype, expecting each endpoint to match a single URL. In preliminary work, we also detect event interaction through message brokers, typical for cloud-native~systems.

% Static analysis can target concerns scattered across microservices to provide a focal point and overview, which is missing in cloud-native systems. Still, more work has to be done to enable support for broad range of platforms or strengthen ingredients to interweave microservice IRs.

% what has to be done

% intermediate model

% - we do consider usage of components to simplify it

% platform-specific parsing

% overlap identification

% - hoslitic overview

\section{Discussion on Implications}

Using our prototype tools, we managed to assess benefits, limitations, and implications from static analysis over cloud-native system with broader detail. 

The primary motivation behind the static analysis is automated reasoning and reports. With the ability to operate across the holistic system or across multiple microservices, developers (as opposed to DevOps) gain new aid to understanding the impact of their changes. For instance, if we track interconnection that disappears with an update, something might be wrong with the update. This would greatly help conformance/consistency checking, which is currently very fragile due to horizontal separation of duty where distinct development teams manage different microservice codebases. However, specific strategies to do so remain to be addressed.

It becomes easier to assess whether the system complies with various organizational policies with holistic system IR. 
Analysts might need to assess the codebase to determine compliance. Similarly to consistency checking, certain policies could be assessed. One venue for discussion and research is the consistency of business logic. Analyzing business logic is difficult from code, even though we know that service components are to be encapsulated and we can track control and data flows. This opens the question of consistency checking across microservices which is certainly attractive. This, however, has to take into account rules added by method interception. Frameworks like Drools can greatly reduce management efforts for business rules; however, these again apply to a single codebase. An integration with configuration server in cloud-native methodology could open a path for centralizing rules.

Another important avenue for research is the centralization of concerns that are now scattered in the decentralized codebases of microservices. 
% In fact, what we discussed above can be generalized. 
For instance, motivation to access the context map or canonical data model, or to have a single focal point for security assessment, understand all business constraints applied in the system. All these perspectives are needed to make wise design decisions, and all these are very difficult to gain from cloud-native systems.

Despite the above details, we need visual representation to properly articulate the concerns, point out the consistency, and sketch the system architectural perspective. A lot of work has been done in this context, recognizing that architecture can be described through various views~\cite{techRep}. The process is known as Software Architecture Reconstruction (SAR)~\cite{10.1007/978-3-030-49418-6_21} can indeed be accomplished through static analysis~\cite{walker2021automatic}. Many researchers address this for microservices either through manual code and documentation assessment~\cite{10.1007/978-3-030-49418-6_21} or through dynamic analysis, which provides only a subset of what is necessary. SAR fits well to the process we described, with the reasoning being recognized as one outcome and visual representation~as~another. 

% \begin{wrapfigure}{r}{0.530\textwidth} 
% %\begin{figure}[h]
% \vspace{-1.3em}
% \hspace{-0.3em}
%\begin{figure}[h]
\begin{figure}
    \centering
       \includegraphics[width=22em]{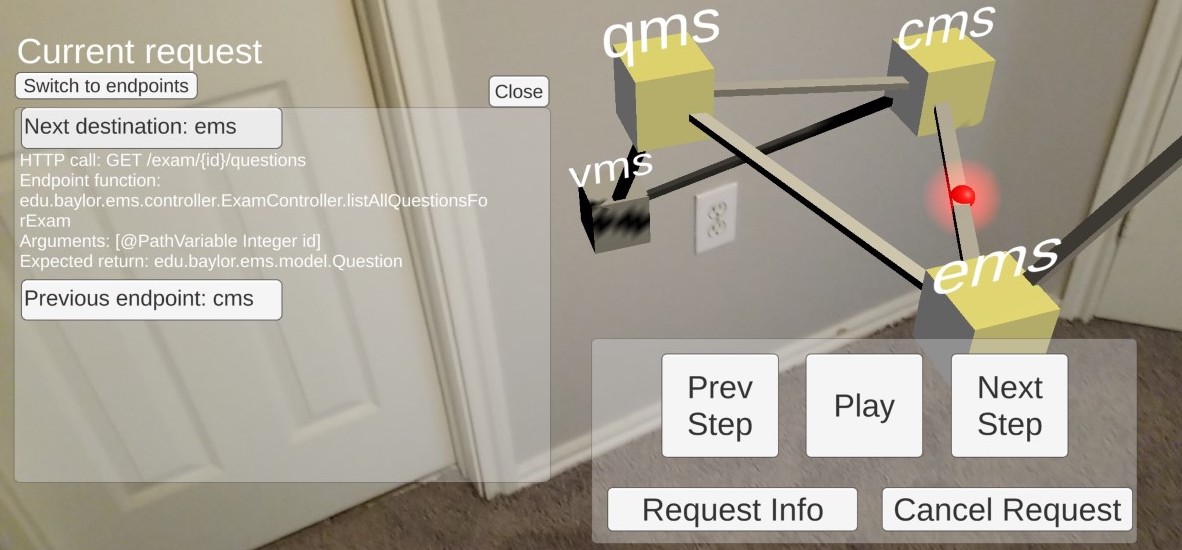}
       \vspace{-3mm}
  \caption{Our AR prototype for SAR and communication simulation.}
%   \,(red\,ball\,moves via\,modules)}
\vspace{-3mm}
  \label{fig:my_label}
\end{figure}

%   \includegraphics[width=1\linewidth]{./img/AR1.png}
% \vspace{-2.1em}

% \label{fig:AR}
% \vspace{-1.3em}
% \end{wrapfigure}

We have researched SAR in cloud-native and used reasoning to detect access policy consistency errors~\cite{das2021automated} or to detect bad smells~\cite{walker2020smells}. We have also researched visualization of microservices~\cite{closer2022vincent} with work in progress on three-dimensional rendering using augmented reality
 as sketches in Figure \ref{fig:my_label} 
for one of the systems testbeds.

A large set of challenges was identified by practitioners in a large study on microservice evolvability~\cite{Bogner2021}. 
% Among many 
Missing system-centric perspectives, inter-service dependencies, coordination between decentralized teams, and challenges with outdated documentation, were the most frequently mentioned issues. They also mention challenges related to microservice integration, API breaking changes, etc. We strongly believe all these could be leveraged by introducing robust static analysis tools for cloud-native systems. 

This article does not anticipate that static analysis is superior to dynamic analysis. It is simply meant for different goals and challenges, and clearly, a broad research opportunity exists for combined static and dynamic analysis.

% bogner

% Reasoning
% - policy compliance 
% - conformance/consistency checking - fragile to overlap identification
% - concerns

% SAR
% - Assessment
% - Analysis

% Integration with dynamic analysis

\section{Conclusion}

In this article, we discuss static analysis in the context of cloud-native design. While static analysis is recognized for many benefits, it has not been widely adopted and used for challenges faced in cloud-native systems. We have listed major obstacles preventing static analysis from operating on the holistic system and point to our experiments that attempted to interweave intermediate representations of microservices to enable such operation. We believe the scientific and industrial community should put more effort into developing robust tools that can help developers better face system evolution and maintenance tasks. In future work, we plan to continue our research in this direction of combining decentralized systems. We will also continue to develop prototypes across languages demonstrating the ability to assess heterogeneous systems to provide a single focal point when assessing certain information and concerns.

%------------------------------------------------------------------------------
\section{Acknowledgments}
\label{sect:acks}
\vspace{-.5em}
This material is based upon work supported by the National Science Foundation under Grant No. 1854049, a grant from  Red Hat Research \url{https://research.redhat.com}, and the ADOMS Grant from Ulla Tuominen Foundation.
\vspace{-.5em}

\label{sect:bib}
\bibliographystyle{plain}
\bibliography{easychair,clowee}

\begin{thebibliography}{10}

\bibitem{Bogner2021}
J.~Bogner, J.~Fritzsch, S.~Wagner, and A.~Zimmermann.
\newblock Industry practices and challenges for the evolvability assurance of
  microservices.
\newblock {\em Empirical Software Engineering}, 26(5):104, 2021.

\bibitem{closer2022vincent}
V.~Bushong, D.~Das, and T.~Cerny.
\newblock Reconstructing the holistic architecture of microservice systems
  using static analysis.
\newblock In {\em Int. Conf. on Cloud Computing and Services Science (CLOSER)},
  2022.

\bibitem{carnell2021spring}
John Carnell and Illary~Huaylupo S{\'a}nchez.
\newblock {\em Spring microservices in action}.
\newblock Manning Publications Co., 2nd ed. Shelter Island, NY, USA, 2021.

\bibitem{testbed}
Tomas Cerny.
\newblock {Microservice Testbed for Texas Teacher Examination}, 2020.
\newblock https://github.com/cloudhubs/tms2020, last accessed 1/2/2022.

\bibitem{CernyIEEEAccess2020}
Tomas Cerny, Jan Svacina, Dipta Das, Vincent Bushong, Miroslav Bures, Pavel
  Tisnovsky, Karel Frajtak, Dongwan Shin, and Jun Huang.
\newblock On code analysis opportunities and challenges for enterprise systems
  and microservices.
\newblock {\em IEEE Access}, 8:159449--159470, 2020.

\bibitem{das2021automated}
D.~Das, A.~Walker, V.~Bushong, J.~Svacina, T.~Cerny, and V.~Matyas.
\newblock On automated rbac assessment by constructing a centralized
  perspective for microservice mesh.
\newblock {\em PeerJ Computer Science}, 7:e376, 2021.

\bibitem{Esparrachiari:2018:TCM:3277539.3277541}
Silvia Esparrachiari, Tanya Reilly, and Ashleigh Rentz.
\newblock Tracking and controlling microservice dependencies.
\newblock {\em Queue}, 16(4):10:44--10:65, August 2018.

\bibitem{DeathStarBench}
Y.~Gan, Y.~Zhang, D.~Cheng, A.~Shetty, P.~Rathi, N.~Katarki, A.~Bruno, J.~Hu,
  B.~Ritchken, B.~Jackson, K.~Hu, M.~Pancholi, Y.~He, B.~Clancy, C.~Colen,
  F.~Wen, C.~Leung, S.~Wang, L.~Zaruvinsky, M.~Espinosa, R.~Lin, Z.~Liu,
  J.~Padilla, and C.~Delimitrou.
\newblock An open-source benchmark suite for microservices and their
  hardware-software implications for cloud \& edge systems.
\newblock In {\em Int. Conf. on Architectural Support for Programming Languages
  and Operating Systems}, 2019.

\bibitem{recovering_architecture}
G.~{Granchelli}, M.~{Cardarelli}, P.~D. {Francesco}, I.~{Malavolta},
  L.~{Iovino}, and A.~D. {Salle}.
\newblock Towards recovering the software architecture of microservice-based
  systems.
\newblock In {\em 2017 IEEE International Conference on Software Architecture
  Workshops (ICSAW)}, pages 46--53, 2017.

\bibitem{han-etal-2013-umbc}
L.~Han, A.~L.~Kashyap, T.~Finin, J.~Mayfield, and J.~Weese.
\newblock {UMBC}{\_}{EBIQUITY}-{CORE}: Semantic textual similarity systems.
\newblock In {\em Conf. on Lexical and Computational Semantics}, 2013.

\bibitem{attack_graph}
A.~Ibrahim, S.~Bozhinoski, and A.~Pretschner.
\newblock Attack graph generation for microservice architecture.
\newblock In {\em Symposium on Applied Computing}, 2019.

\bibitem{Imranur2019}
{M}ohammad~{R}ahman Imranur, {S}ebastiano Panichella, and {D}avide Taibi.
\newblock A curated dataset of microservices-based systems.
\newblock In {\em Joint Proceedings of the Summer School on Software
  Maintenance and Evolution}. CEUR-WS, 09 2019.

\bibitem{service_dependency}
S.~{Ma}, C.~{Fan}, Y.~{Chuang}, W.~{Lee}, S.~{Lee}, and N.~{Hsueh}.
\newblock Using service dependency graph to analyze and test microservices.
\newblock In {\em 42nd Annual Computer Software and Applications Conf.}, 2018.

\bibitem{techRep}
Liam O'Brien, Christoph Stoermer, and Chris Verhoef.
\newblock Software architecture reconstruction: Practice needs and current
  approaches.
\newblock Technical report, Carnegie Mellon University, 01 2002.

\bibitem{Panichella2021}
{S}ebastiano Panichella, {M}ohammad~{R}ahman Imranur, and {D}avide Taibi.
\newblock Structural coupling for microservices.
\newblock In {\em 11th International Conference on Cloud Computing and Services
  Science}, 04 2021.

\bibitem{Pigazzini2020}
{I}laria Pigazzini, {F}rancesca Arcelli~Fontana, {V}alentina Lenarduzzi, and
  {D}avide Taibi.
\newblock Towards microservice smells detection.
\newblock In {\em Proceedings of the 3rd International Conference on Technical
  Debt}, TechDebt '20, page 92–97, New York, NY, USA, 2020.

\bibitem{10.1007/978-3-030-49418-6_21}
F.~Rademacher, S.~Sachweh, and A.~Z{\"u}ndorf.
\newblock A modeling method for systematic architecture reconstruction of
  microservice-based software systems.
\newblock In {\em Enterprise, Business-Process and Information Systems
  Modeling}. Springer International Publishing, 2020.

\bibitem{Schiewe2022}
Micah Schiewe, Jacob Curtis, Vincent Bushong, and Tomas Cerny.
\newblock Advancing static code analysis with language-agnostic component
  identification.
\newblock {\em IEEE Access}, 10:30743--30761, 2022.

\bibitem{spe.2974}
J.~Soldani, G.~Muntoni, D.~Neri, and A.~Brogi.
\newblock The ntosca toolchain: Mining, analyzing, and refactoring
  microservice-based architectures.
\newblock {\em Sw. Practice and Experience}, 51(7):1591--1621, 2021.

\bibitem{Taibi2019closer19}
{D}avide Taibi and {K}ari Syst{\"a}.
\newblock From monolithic systems to microservices: A decomposition framework
  based on process mining.
\newblock In {\em Proceedings of the 9th International Conference on Cloud
  Computing and Services Science - Volume 1: CLOSER,}, pages 153--164. INSTICC,
  SciTePress, 2019.

\bibitem{Taibi2020CLOSER}
{D}avide Taibi and {K}ari Syst{\"a}.
\newblock A decomposition and metric-based evaluation framework for
  microservices.
\newblock In {\em Cloud Computing and Services Science}, pages 133--149, Cham,
  2020. Springer International Publishing.

\bibitem{Thalheim:2017:SAI:3135974.3135977}
J~Thalheim, A~Rodrigues, I.E. Akkus, P~Bhatotia, Rm~Chen, B.~Viswanath,
  L.~Jiao, and C.~Fetzer.
\newblock Sieve: Actionable insights from monitored metrics in distributed
  systems.
\newblock In {\em Middleware}, 2017.

\bibitem{walker2020smells}
A.~Walker, D.~Das, and T.~Cerny.
\newblock Automated code-smell detection in microservices through static
  analysis: A case study.
\newblock {\em Applied Sciences}, 10(21), 2020.

\bibitem{walker2021automatic}
A.~Walker, I.~Laird, and T.~Cerny.
\newblock On automatic software architecture reconstruction of microservice
  applications.
\newblock {\em Information Science and Applications}, 2021.

\bibitem{trainticket}
X.~Zhou, X.~Peng, T.~Xie, J.~Sun, C.~Xu, C.~Ji, and W.~Zhao.
\newblock Benchmarking microservice systems for software engineering research.
\newblock In {\em 40th Int. Conf.Software Engineering: Comp.}, 2018.

\end{thebibliography}

%----------------------------------------------------------------------
\end{document}